%Paper: hep-th/9206080
%From: MUKHI%tifrvax.BITNET@pucc.princeton.edu
%Date: Sun, 21 Jun 92 13:28 IST +0530

%%%%%%%%%%%%%%%%%%%%%%%%%%%%%%%%%%%%%%%%%%%%%%%%%%%%%%%%%%%%%%%%%%%%%%

%%%%%%  PROCESS USING THE TEX MACRO "PHYZZX"  %%%%%%%%%%%%%%%%%%%%%%%%

%%%%%%%%%%%%%%%%%%%%%%%%%%%%%%%%%%%%%%%%%%%%%%%%%%%%%%%%%%%%%%%%%%%%%%

\input phyzzx
\NPrefs
\def\define#1#2\par{\def#1{\Ref#1{#2}\edef#1{\noexpand\refmark{#1}}}}
\def\con#1#2\noc{\let\?=\Ref\let\<=\refmark\let\Ref=\REFS
         \let\refmark=\undefined#1\let\Ref=\REFSCON#2
         \let\Ref=\?\let\refmark=\<\refsend}

\define\POLYAKOV
A.M. Polyakov, Mod. Phys. Lett. {\bf A6} (1991) 635.

\define\LZ
B. Lian and G. Zuckerman, Phys. Lett. {\bf B266} (1991) 21.

\define\MMS
S. Mukherji, S. Mukhi and A. Sen, Phys. Lett. {\bf B266} (1991) 337.

\define\BMP
P. Bouwknegt, J. McCarthy and K. Pilch, Comm. Math. Phys. {\bf 145} (1992)
541.

\define\WITTEN
E. Witten, Nucl. Phys. {\bf B373} (1992) 187.

\define\WITZWIE
E. Witten and B. Zwiebach, Princeton-MIT preprint IASSNS-HEP-92/4,
MIT-CTP-2057 (January 1992).

\define\IM
C. Imbimbo and S. Mukhi, Nucl. Phys. {\bf B364} (1991) 662.

\define\GINSPARG
P. Ginsparg, Nucl. Phys. {\bf B295} (1988) 153.

\define\HARRIS
G. Harris, Nucl. Phys. {\bf B300} (1988) 588.

\define\DVV
R. Dijkgraaf, E. Verlinde and H. Verlinde, Comm. Math. Phys. {\bf 115}
(1988) 649.

\define\VERLINDE
E. Verlinde, Princeton preprint IASSNS-HEP-92/5 (February 1992).

\define\SLODOWY
P. Slodowy, in Lecture Notes in Mathematics, 1008, ed. J. Dolgachev
(Springer-Verlag, Berlin, 1983);\hfill \break
P. Slodowy, ``Simple Singularities and Simple Algebraic Groups'', Lecture
Notes in Mathematics, 815 (Springer-Verlag, Berlin, 1980).

\define\KLEIN
F. Klein, ``Lectures on the Icosahedron, and the Solution of Equations of
the Fifth Degree'' (English Translation) (Dover, New York, 1956).

\define\BARBON
J. Barb\'on, CERN preprint CERN-TH.6379/92 (January 1992).

\define\KACHRU
S. Kachru, Princeton preprint PUPT-1305 (January 1992).

\define\EGUCHI
T. Eguchi and A. Hanson, Phys. Lett. {\bf B74} (1978) 249; Ann. Phys.
{\bf 120} (1979) 82.

\define\GIBBONS
G. Gibbons and S. Hawking, Phys. Lett. {\bf B78} (1978) 430.

\define\HAWKING
S. Hawking, in Recent Developments in Gravitation, Carg\`ese 1978, ed. M.
L\'evy and S. Deser (Plenum Press).

\define\HITCHIN
N. Hitchin, Math. Proc. Camb. Phil. Soc. {\bf 85} (1979) 465.

\define\KRONHEIMER
P. Kronheimer, J. Diff. Geom. {\bf 29} (1989) 665, 685.

\define\OOGURIVAFA
H. Ooguri and C. Vafa, Mod. Phys. Lett. {\bf A5} (1990) 1389; Nucl. Phys.
{\bf B361} (1991) 469.

\define\GIVEON
A. Giveon and A. Shapere, Cornell-Princeton preprint CLNS-92/1139,
IASSNS-HEP-92/14 (February 1992).

\define\TAKASAKI
K. Takasaki, Kyoto preprint KUCP-0049/92 (June 1992).

\define\KLEBAPOLY
I. Klebanov and A. Polyakov, Mod. Phys. Lett. {\bf A6} (1991) 3273.

\define\DANCER
A. Dancer, DAMTP preprint 92-13, 92-19.

\define\PLEBANSKI
J. Plebanski, J. Math. Phys. {\bf 16} (1975) 2395.

\font\sans=cmssbx10 scaled\magstep1

\def\half{{1\over2}}
\def\bz{{\bar z}}
\def\Ox{{\cal O}_{\half,\half}}
\def\Oy{{\cal O}_{\half,-\half}}
\def\Osn{{\cal O}_{s,n}}

\def\bOsn{\overline{{\cal O}}_{s,n'}}
\def\bOsmn{\overline{{\cal O}}_{s-1,n'}}
\def\del{\partial}
\def\bx{{\bar x}}
\def\by{{\bar y}}
\def\bY{\overline{Y}}
\def\bb{{\bar b}}
\def\ba{{\bar a}}
\def\bh{{\bar h}}
\def\CC{{\sans C}}
\def\RR{{\sans R}}
\def\cyc{{\cal C}}
\def\dihed{{\cal D}}
\def\tet{{\cal T}}
\def\oct{{\cal O}}
\def\icos{{\cal I}}
\def\del{\partial}
\def\delbar{{\bar\partial}}
\def\delx{{\partial\over\partial x}}
\def\dely{{\partial\over\partial y}}
\def\delaone{{\partial\over\partial a_1}}
\def\delatwo{{\partial\over\partial a_2}}
\def\delathree{{\partial\over\partial a_3}}
\def\delafour{{\partial\over\partial a_4}}
\def\delai{{\partial\over\partial a_i}}
\def\tW{{\tilde W}}
\def\tZ{{\tilde Z}}
\def\delX{{\partial\over\partial X}}
\def\delY{{\partial\over\partial Y}}
\def\delZ{{\partial\over\partial Z}}
\def\delW{{\partial\over\partial W}}
\def\deltZ{{\partial\over\partial \tZ}}
\def\deltW{{\partial\over\partial \tW}}
\def\to{\rightarrow}

{\nopagenumbers

\ \hfill\vbox{\hbox{TIFR/TH/92-34}\hbox{June, 1992}}\break

\title{KLEINIAN SINGULARITIES AND THE GROUND RING OF C=1 STRING THEORY}

\author{Debashis Ghoshal, Dileep P. Jatkar and Sunil Mukhi}\foot{e-mail:
ghoshal@tifrvax.bitnet, dileep@tifrvax.bitnet, mukhi@tifrvax.bitnet}

\address{Tata Institute of Fundamental Research \break Homi Bhabha Road,
Bombay 400005, India}

\abstract

We investigate the nature of the ground ring of $c=1$ string theory at the
special $A$-$D$-$E$ points in the $c=1$ moduli space associated to discrete
subgroups of $SU(2)$. The chiral ground rings at these points are shown to
define the $A$-$D$-$E$ series of singular varieties introduced by Klein. The
non-chiral ground rings relevant to closed-string theory are 3 real
dimensional singular varieties obtained as $U(1)$ quotients of the
Kleinian varieties. The unbroken symmetries of the theory at these points
are the volume-preserving diffeomorphisms of these varieties.
The theory of Kleinian singularities has a close relation
to that of complex hyperK\"ahler surfaces, or gravitational instantons. We
speculate on the relevance of these instantons and of self-dual gravity in
$c=1$ string theory.

\endpage }

\section{\bf Introduction}

An important advance in the understanding of non-critical $c=1$ string
theory was the observation\POLYAKOV\ that a large class of the physical
states in this theory may be identified with the residual modes, in
two spacetime dimensions, of the graviton and higher-spin gauge fields
that are familiar from the study of critical strings. These states occur
for special values of the matter momentum. In addition to these ``discrete
states'', which arise due to the presence of null vectors in the matter
CFT, there are other tachyon-like states whose spectrum in the
infinite-radius theory is continuous.

{}From this point of view, it seems worthwhile to use $c=1$ string theory as
a toy model to understand the enormous variety of physical states that
appear in the critical case, and to investigate the symmetries associated
to them.

A seemingly unrelated development was the discovery\con\LZ\MMS\BMP\noc\
that the BRS cohomology in the $c=1$ background has infinitely many extra
states of non-standard ghost number, a fact which has essentially no
counterpart in critical string theory. A powerful approach to the
understanding of these states was presented in Ref.\WITTEN, where it was
noted that a subclass of them forms a polynomial ring (the ``ground
ring'') which controls the symmetry properties of the theory. It was shown
that the symmetries are associated to volume-preserving diffeomorphisms of
the variety defined by the polynomial ground ring. One curious consequence
of this approach is that this variety, which is three-dimensional (or
four-dimensional in a certain sense) for many of the possible $c=1$
backgrounds, seems to play the role of a target spacetime in $c=1$ string
theory, even though conventionally the target spacetime in this theory is
thought to be two-dimensional.

The study of ground rings at $c=1$ was taken further in\WITZWIE, where in
particular the volume-preserving symmetries were related to the presence
of string field theory gauge parameters, of ghost number 1, which are in
the BRS cohomology. (Precisely the same suggestion, in the context of
$c<1$ backgrounds, had been made earlier in Ref.\IM, where however it
could not be made more explicit due to the unavailability of a complete
classification of the cohomology at the time.) It was also shown in
Ref.\WITZWIE\ that the closed string theory has even more ``extra states''
in the cohomology than was previously thought.

Virtually all of the analysis above was done in the closed string theory
where the matter coordinate is either noncompact or compactified at the
self-dual radius. In the latter case, the matter theory acquires an
$SU(2)\otimes SU(2)$ symmetry via the Frenkel-Kac-Segal mechanism, and
this is evidently the phase of the string theory with the maximum number
of unbroken symmetries.

It is known\con\GINSPARG\HARRIS\noc\ (see also \DVV) that the moduli space
of $c=1$ conformal field theory has a very nontrivial and elegant
structure, including two critical lines (associated to circle and
$Z_2$-orbifold compactifications of the free scalar fields) intersecting
at a multicritical point, and three isolated points disconnected from the
above lines. It is of interest to understand how the symmetries of $c=1$
string theory change as we move in the $c=1$ CFT moduli space. This is
important particularly because we need to understand how symmetry-breaking
takes place in string theory. (One approach to this question was proposed
recently in Ref.\VERLINDE.)

It is reasonably straightforward to construct the ground ring and
symmetries explicitly at all points of the CFT moduli space, as we will
show below. What is remarkable is that, for a special subset of points in
this moduli space, the result of this investigation turns out to have a
beautiful connection with the theory (studied by Klein\KLEIN\ in the last
century) of polynomials on \CC$^2$ invariant under discrete subgroups of
$SU(2)$, and the associated algebraic varieties. This theory in turn has,
in the last two decades, proved to be intimately related to self-dual
gravity in four spacetime dimensions%
\con\EGUCHI\GIBBONS\HAWKING\HITCHIN\KRONHEIMER\DANCER\TAKASAKI\noc. This
therefore suggests a crucial role for self-dual gravity in $c=1$ string
theory (which has been suggested before on other
grounds\con\OOGURIVAFA\GIVEON\noc).

In what follows, we first describe the explicit construction of the chiral
ground ring of $c=1$ string theory at a set of special ($A$-$D$-$E$)
points in the moduli space, and show that the corresponding spaces are
precisely the Kleinian singular varieties. Next we combine left and right
movers in closed string theory and obtain the full non-chiral ground rings
(called ``quantum ground rings'' in Ref.\WITTEN). These are shown to be
certain quotients of the Kleinian varieties. For completeness, we also
present a classification of polynomial ground rings of $c=1$ string theory
at other points in the moduli space, where they are not related to
Kleinian varieties.  Finally, we consider the four-dimensional topological
action obtained in Ref.\WITZWIE\ at the $SU(2)$ point, and argue that it
has nontrivial solutions at the other $A$-$D$-$E$ points, corresponding to
hyperK\"ahler surfaces or gravitational instantons. It is here that the
connection to self-dual gravity emerges.

\section{\bf The Ground Ring of $c=1$ String Theory}

The BRS cohomology of $c=1$ string theory contains, in particular,
operators of zero ghost number\LZ\MMS\BMP. It was noted in
Ref.\WITTEN\ that these form a ring under the operator product, because
ghost number 0 is obviously invariant under this product, and there are no
singularities in the OPE since cohomology elements all have conformal
dimension 0.

The structure of this ring was analyzed in detail\WITTEN\WITZWIE\
for the case where the matter field $X(z,\bz)$ is compactified on a circle
of radius $1\over\sqrt2$, the self-dual or $SU(2)$ point. In this case,
the chiral ground ring, consisting of the purely holomorphic part of the
(relative) cohomology elements, is the free ring on two generators
$$
\eqalign{
x &\equiv \Ox(z) \equiv ~:\left(c(z)b(z) + i\del X^-(z)\right)
e^{i X^+(z)}: \cr
y &\equiv \Oy(z) \equiv ~:\left(c(z)b(z) - i\del X^+(z)\right)
e^{-i X^-(z)}: \cr}
\eqn\one
$$
where $X^\pm\equiv {1\over\sqrt2}(X\pm i\phi)$ and $\phi$ is the Liouville
field.

This ring should be relevant for open-string theory, while in
closed-string theory we need to combine the holomorphic pieces with their
anti-holomorphic counterparts, keeping in mind that the Liouville momentum
in the left and right sectors must be equal since the Liouville field is
non-compact. It follows that, if $\bx$ and $\by$ are the anti-holomorphic
counterparts of $x$ and $y$ defined above, the non-chiral ground ring is
generated by the four operators
$$
a_1\equiv x\bx,\quad a_2\equiv y\by,\quad a_3\equiv
x\by,\quad a_4\equiv y\bx
\eqn\onea
$$
satisfying the relation
$$
a_1 a_2 = a_3 a_4
\eqn\two
$$
Thus we have a ring on four generators satisfying one relation. The
set of ghost-number 0 operators obtained by taking powers of the
generators above is given by
$$
\Osn(z)\bOsn(\bz)
\eqn\fivea
$$
where
$$
\Osn(z)~\sim~ h_{s,n}\equiv x^{s+n} y^{s-n}
\eqn\new
$$
and similarly for the antiholomorphic part. Here, $s$ and $n$ are
half-integers labelling the total isospin and $J_3$-eigenvalue with
respect to $SU(2)$.

In the language of algebraic geometry, a polynomial ring defines a
variety. We will take the point of view that the chiral ring defines a
complex affine variety in \CC$^2$ (in the present case it is just \CC$^2$
itself), while the non-chiral ring defines a real affine variety in
\RR$^4$ (in the present case it is a quadric cone). This is natural as the
chiral ring is associated to holomorphic conformal fields, while the
non-chiral ring is associated to real fields.

It has been shown in \WITTEN\WITZWIE\ that the existence of the ground
ring in $c=1$ string theory is responsible for the existence of a large
algebra of unbroken symmetries. This comes about because symmetry
generators in closed-string theory are associated to states of ghost
number 1 which are annihilated by the antighost zero mode $b_0^-\equiv b_0
- \bb_0$. These are obtained by combining holomorphic states of ghost
number 1 (the chiral version of tachyons and discrete states) with
anti-holomorphic states of ghost number 0 (the ground ring elements) and
vice versa. At the $SU(2)$ point, the chiral ground ring elements are
given by the $\Osn(z)$ of Eq.\new,
while the relevant chiral operators of ghost number 1 are
$$
Y^+_{s,n} \equiv ~:c(z) V_{s,n}(z)e^{\sqrt2(1-s)\phi(z)}:
\eqn\four
$$
where $V_{s,n}$ are the $SU(2)$ multiplets which make up the primary fields
in the matter sector. Thus, the symmetry generators are
$$
Y^+_{s,n}(z)
\bOsmn(\bz)
\eqn\five
$$
and the conjugate ones.

In addition, there is a set of ghost-number 1 operators whose existence
follows from the fact that $a + \ba \equiv [Q_B, \phi]$ is in the
cohomology if we restrict to the set of usual conformal fields. This leads
to the ``new'' symmetry generators
$$
:\left(a(z) + \ba(\bz)\right)\Osn(z)\bOsn(\bz):
\eqn\fiveaa
$$

In what follows, we will investigate the nature of the ground ring of
$c=1$ string theory at various points in the moduli space of $c=1$
conformal field theories.

\section{\bf $A$-$D$-$E$ Classification of Chiral Ground Rings}

The conformal field theory of a single free boson has a special set of
points in its moduli space associated to quotients of $SU(2)$ by its
discrete subgroups, which are the binary groups classified as cyclic
(${\cal C}_{n}$), dihedral (${\cal D}_n$), tetrahedral (${\cal T}$),
octahedral (${\cal O}$) and icosahedral (${\cal I}$)\GINSPARG\HARRIS.
These finite groups are in turn associated with the Dynkin diagrams of the
$A,D$ and $E$ series of simply-laced Lie algebras\SLODOWY. The special
points in the moduli space correspond to starting with the $c=1$ theory
compactified on a circle of self-dual radius $R={1\over\sqrt2}$, where the
theory has an $SU(2)\otimes SU(2)$ symmetry, and taking the quotient of
$SU(2)$ by its various discrete subgroups. The cyclic or $A_n$-type
subgroups lead to theories of a free boson compactified on a circle of
radius $n/\sqrt2$, while the dihedral or $D_n$-type subgroups give
theories of a free boson compactified on a $Z_2$-orbifold of the circle,
at radius $n/\sqrt2$ (in both these cases, replacing $n$ by $1/n$ gives
the same theory, by duality). The remaining three discrete subgroups,
associated to the Dynkin diagrams of $E_6,E_7$ and $E_8$, lead to a set of
three models which have no integrable moduli and are disconnected (in the
$c=1$ CFT) from the circle and orbifold compactifications\GINSPARG\HARRIS.

To find the ground rings at these points, we need to keep those
generators of the $SU(2)$ ground ring that survive the modding out
procedure. It is clear that the basic generators $x,y$ defined in Eq.\one\
form a doublet under $SU(2)$, since they are obtained from each other by
the action of the $SU(2)$ raising and lowering operators $J^\pm$ of the
$c=1$ CFT. Thus the elements of the ground ring at the $A$-$D$-$E$ points
are obtained as all polynomials in $x,y$ which are invariant under the
corresponding discrete group $\Gamma$ acting as a subgroup of $SU(2)$.

The mathematical problem of characterising the polynomials in two complex
variables which are invariant under any discrete subgroup $\Gamma$ of
$SU(2)$ was addressed and solved in the last century by Klein\KLEIN. To
each binary subgroup $\Gamma$, Klein associated a polynomial ring in three
variables $X,Y$ and $Z$, with one relation between them. This defines a
complex affine variety in \CC$^3$, which is smooth everywhere
except at the origin, where it is singular in general. These are commonly
known as Kleinian singular varieties or simply Kleinian singularities.
Using this result of Klein, it is an easy matter to describe the chiral
ground rings at each of the special points in the moduli space of $c=1$
conformal field theory. Note that the problem of finding the {\it
non-chiral} ground ring is more complicated and we will discuss it in some
detail in the following section.

Let us start with the the $A$-series. The corresponding discrete subgroups
of $SU(2)$ are the binary cyclic groups $\cyc_n= Z_{2n}$, whose action
on $x,y$ is generated by
$$
\pmatrix{x\cr y\cr} \rightarrow \pmatrix{e^{\displaystyle i\pi/n}& 0\cr
0 & e^{\displaystyle -i\pi/n}\cr} \pmatrix{x\cr y\cr}
\eqn\six
$$
The smallest set which generates all polynomials invariant under $\cyc_n$
is clearly
$$
X\equiv x^{2n},\quad Y\equiv y^{2n},\quad Z\equiv xy
\eqn\seven
$$
and these three generators obey the single relation
$$
XY = Z^{2n}
\eqn\eight
$$
This defines the chiral ground ring at the $R=n/\sqrt2$ points, as a
variety in \CC$^3$.

At this point there is a slight puzzle, since the case $n=1$ in the above
equation gives a quadric in \CC$^3$, rather than just \CC$^2$ as was found
in Ref.\WITTEN\ for the chiral ground ring at the $SU(2)$ point. This
really has to do with the meaning of chiral ground ring. As we noted
earlier, this ring is just an auxiliary construction for closed-string
theory, which acquires a direct physical interpretation only after
combining left and right movers. In this process, the transformation
$(x,y)\rightarrow -(x,y)$ is an invariance, from Eq.\onea. Thus if we are
interested in constructing the building blocks from which to make the
non-chiral ground ring, then we need not impose this $Z_2$ invariance
separately on the chiral ring elements. This means that we should quotient
not by the binary but rather by the ordinary subgroups of $SU(2)$, which
descend to subgroups of $SO(3)$. We will not take this point of view in
the present work, since it is the chiral ground ring constructed from
binary subgroups which is most closely related to the Kleinian
singularities. In a subsequent section, we will construct the correct
non-chiral ground ring following a procedure analogous to the one above.

Next we look at the $D$-series. The binary dihedral group $\dihed_n$
is the set of transformations generated by the generators of $\cyc_n$
together with
$$
\pmatrix{x\cr y\cr} \rightarrow \pmatrix{0 & i\cr
i & 0\cr} \pmatrix{x\cr y\cr}
\eqn\nine
$$
The generators of invariant polynomials are
$$
X= \half(x^{2n} + (-1)^n y^{2n}), \quad Y= \half xy\;(x^{2n} - (-1)^n y^{2n}),
\quad Z= (xy)^2
\eqn\ten
$$
These obey the relation
$$
Y^2 = Z(X^2 - Z^n)
\eqn\eleven
$$
which therefore defines the chiral ground ring at the orbifold points at
radius $R=n/\sqrt2$.

Finally, for the exceptional points we build up the chiral ground ring as
follows. For $E_6$, corresponding to the binary tetrahedral group $\tet$,
the polynomial invariants are obtained by starting with the binary
dihedral group $\dihed_2$ and adjoining the action of the single generator
$$
T = {1\over\sqrt2}\pmatrix{\epsilon^7 & \epsilon^7\cr \epsilon^5 &
\epsilon\cr}
\eqn\twelve
$$
where $\epsilon$ is a primitive eighth root of unity.
It is easy to deduce that the invariants of $\tet$ are given by suitable
combinations of $\dihed_2$ invariants, whose explicit expressions in terms
of $x$ and $y$ are
$$
\eqalign{
X &= {1\over 4}(x^8 + y^8 + 14\; x^4 y^4) \cr
Y &= \half xy\;(x^4 - y^4)\cr
Z &= {1\over 8} (x^4 + y^4)(x^8 + y^8 - 34\; x^4 y^4)\cr   }
\eqn\thirteen
$$
These generators obey the relation
$$
Z^2 = X^3 - 27\; Y^4
\eqn\fourteen
$$
which defines the chiral $E_6$ ground ring.

To get the binary octahedral group $\oct$ one must adjoin the matrix
$$
O = \pmatrix{\epsilon & 0\cr 0 & \epsilon^7\cr}
\eqn\fifteen
$$
to the generators of \tet, where $\epsilon$ is again a primitive eighth
root of unity. The ring elements invariant under this matrix as well are
generated by
$$
\eqalign{
X &= {1\over4}(x^8 + y^8 + 14\; x^4 y^4)\cr
Y &= {1\over4}x^2y^2(x^4-y^4)^2\cr
Z &= {1\over16}xy\; (x^8 - y^8)(x^8 + y^8 - 34\; x^4 y^4) }
\eqn\sixteen
$$
with the relation
$$
Z^2 = Y(X^3 - 27\;Y^2)
\eqn\seventeen
$$

Finally, for the binary icosahedral group \icos, related to $E_8$, we need
to go back to the ground ring at the $SU(2)$ point, generated by $x$ and
$y$, and impose invariance under the two matrices
$$
I_1 = \pmatrix{-\eta^3 &0\cr 0& -\eta^2 \cr} \qquad
I_2 = {1\over \eta^2 - \eta^3}\pmatrix{\eta + \eta^4 & 1\cr
1 & -(\eta + \eta^4)\cr}
\eqn\eighteen
$$
Technically this is the most complicated case to analyse. It turns out
that the three basic polynomials invariant under this set of
transformations are
$$
\eqalign{
X &= xy\;(x^{10} + 11\; x^5 y^5 - y^{10})\cr
Y &= -(x^{20} + y^{20}) + 228\; x^5 y^5(x^{10} - y^{10}) -
494\; x^{10} y^{10}\cr
Z &= (x^{30} + y^{30}) + 522\; x^5 y^5(x^{20} - y^{20}) -
10005\; x^{10} y^{10} (x^{10} + y^{10})\cr   }
\eqn\nineteen
$$
and they satisfy the relation
$$
Z^2 = - Y^3 + 1728\; X^5
\eqn\twenty
$$

This completes the classification of chiral ground rings in $c=1$ string
theory, in terms of the basic generators $x$ and $y$ at the $SU(2)$ point,
for the $A$-$D$-$E$ series of $c=1$ CFT. In each case, we find that the
ground ring defines a Kleinian singularity, which is a variety in \CC$^3$
with a singularity at the origin.

One may ask what happens for other points in the moduli space of this CFT,
namely, circle and orbifold compactifications where the radius is not an
integer multiple of ${1\over\sqrt2}$. These cases cannot be obtained by
modding out the $SU(2)$ theory by a discrete subgroup, hence they are not
related to the Kleinian singularities, but it is straightforward to
analyse them separately. We will consider them below when we construct the
full non-chiral ground ring at various points in the $c=1$ CFT moduli
space.

\section{\bf $A$-$D$-$E$ Classification of Non-Chiral Ground Rings}

In this section we find the invariant polynomials, and the relations
between them, which characterise the ground ring of closed-string theory.
First we consider the $A$-$D$-$E$ series of special points, as before, and
show that the ground rings and their associated varieties can be thought
of as suitable $U(1)$ quotients of the Kleinian singularities. We
construct these explicitly in all the cases except $E_8$, which is
technically very complicated.

We have noted earlier that if $\Gamma$ is a discrete subgroup of $SU(2)$,
then the chiral ground ring for the $SU(2)/\Gamma$ theory is given by
imposing $\Gamma$-invariance on the polynomials in $x$ and $y$, the
analytic polynomials in \CC$^2$. Consider now the antiholomorphic
generators of the $SU(2)$ ground ring, $\bx$ and $\by$, and let us
associate these with the complex conjugate coordinates of \CC$^2$. To find
the non-chiral ground ring, we must consider polynomials in $x, y, \bx,
\by$ which have the same Liouville momenta in the holomorphic and
antiholomorphic parts\WITTEN, and are in addition $\Gamma$-invariant. The
constraint of Liouville momentum matching amounts to taking a $U(1)$
quotient, where the action is defined by
$$
\pmatrix{x\cr y\cr} \rightarrow e^{\displaystyle i\theta}
\pmatrix{x\cr y\cr}\qquad
\pmatrix{\bx\cr \by\cr} \rightarrow e^{\displaystyle -i\theta}
\pmatrix{\bx\cr \by\cr}
\eqn\twentyone
$$
Requiring polynomials to be invariant under this action is equivalent to
requiring that they be annihilated by the vector field
$$
H = x\del_x + y\del_y - \bx\del_\bx - \by\del_\by
\eqn\twentytwo
$$

Now the quotient by $\Gamma$ must be taken on the space of
$U(1)$-invariant polynomials on \CC$^2$. Clearly these polynomials cannot
be analytic in $x$ and $y$ but rather must be built out of the
combinations defined in Eq.\onea, subject to the relation \two. Formally,
we may say that while the (complexified) chiral ground ring is the
Kleinian variety \CC$^2/\Gamma$, the non-chiral ring is
\CC$^2/(\Gamma\otimes U(1))$, which we view as a real variety in \RR$^4$.

To obtain the $A_n$ series, we note that the polynomials built out of
$x\bx$, $y\by$, $x\by$ and $y\bx$ which are invariant under $\cyc_n$ are
generated by
$$
\eqalign{
W &= \half(x\bx + y\by)\cr
X &= (x\by)^n\cr
Y &= (y\bx)^n\cr
Z &= \half(x\bx - y\by)\cr}
\eqn\twentythree
$$
satisfying the relation
$$
(W^2 - Z^2)^n = XY
\eqn\twentyfour
$$
At this stage we observe that the generator $W$ is present at every
$A$-$D$-$E$ point, since it is in fact invariant under the whole of
$SU(2)$. The other three generators have some similarity with the three
generators of the corresponding chiral ground ring. Finally, the relation
between the four generators reduces to the Kleinian relation when $W$ is
set to zero.

Let us also examine the nature of the singularities of the variety defined
by Eq.\twentyfour\ above. Rewriting this equation in the form
$f(X,Y,Z,W)=0$, one finds that the tangent to the variety at any point is
given by the 4-vector (in \RR$^4$)
$$
(\del_X f,\; \del_Y f,\; \del_Z f,\; \del_W f) =
(Y,\;X,\; 2n\;Z\;(W^2 - Z^2)^{n-1},\; - 2n\; W\; (W^2 - Z^2)^{n-1})
\eqn\twentyfoura
$$
This vector vanishes at the set of points
$$
X=0,\qquad Y=0,\qquad Z = \pm W
\eqn\twentyfourb
$$
which defines two intersecting straight lines in \RR$^4$. Thus, unlike
the chiral ground ring where the corresponding variety is singular only at
the origin, the non-chiral ring at the $A_n$ points defines a variety with
lines of singular points. The only exception is the $SU(2)$ point ($n=1$),
where the tangent vector vanishes only at the origin, from
Eq.\twentyfoura.  At the $D$ and $E$ points to be discussed below, we will
find that the varieties have (real) curves of singular points.

For the $D_n$ series, the invariants are easily seen to be
$$
\eqalign{
W &= \half(x\bx + y\by)\cr
X &= \half\left((x\by)^n + (y\bx)^n\right)\cr
Y &= {1\over4}(x\bx - y\by)\left((x\by)^n - (y\bx)^n\right)\cr
Z &= x\bx y\by\cr}
\eqn\twentyfive
$$
with the relation
$$
Y^2 = (W^2 - Z)(X^2 - Z^n)
\eqn\twentysix
$$
It is easy to check that for $n=1$, this is the same as Eq.\twentyfour\ for
$n=2$, under an invertible polynomial redefinition of the generators. This
corresponds to the equivalence of the binary cyclic group $\cyc_2 = Z_4$
and the binary dihedral group $\dihed_1$. The associated point in the
moduli space is the multicritical point at which the lines of circle and
orbifold compactifications intersect.

The singular points for the $D_n$ series are parametrized by the curves
$$
Y=0,\qquad Z = W^2,\qquad X = \pm W^n
\eqn\twentysixa
$$

The $E_6$ case, as before, is obtained by starting with $\dihed_2$
invariants and modding out by the action of the matrix $T$ in Eq.\twelve.
One finds the invariants
$$
\eqalign{
W &= \half(x\bx + y\by)\cr
X &= {1\over4}\left((x\by)^4 + (y\bx)^4 + 14\; (x\bx y\by)^2 \right)
- x\bx y\by\;(x\bx + y\by)^2 \cr
Y &= {1\over 4}(x\bx - y\by)\left((x\by)^2 - (y\bx)^2\right)\cr
Z &= {1\over 8}\left( (x\by)^2 + (y\bx)^2 \right) \left( (x\by)^4 +
(y\bx)^4 - 34\; (x\bx y\by)^2  + 12\; x\bx y\by (x\bx + y\by)^2 - (x\bx +
y\by)^4 \right)\cr}
\eqn\twentyseven
$$
satisfying the relation
$$
Z^2 = X^3 - 27\;Y^4 + W^2(18\; X Y^2 + X^2 W^2 + 16\; Y^2 W^4)
\eqn\twentyeight
$$
This is the same as Eq.\fourteen\ when $W$ is set to zero.
The singular points can be shown to be parametrized by the curves
$$
Z=0,\qquad X = -{4\over3}\; W^4, \qquad Y = \pm{2i\over 3\sqrt{3}}\; W^3
\eqn\twentyeighta
$$

For $E_7$, we quotient the $E_6$ system by the matrix $O$ in Eq.\fifteen,
and find the invariants
$$
\eqalign{
W =& \half(x\bx + y\by)\cr
X =& {1\over4}\left((x\by)^4 + (y\bx)^4 + 14\; (x\bx y\by)^2 \right)
- x\bx y\by\;(x\bx + y\by)^2 \cr
Y =& {1\over 16} (x\bx - y\by)^2 \left((x\by)^2 - (y\bx)^2\right)^2 \cr
Z =& {1\over32} (x\bx - y\by) \left((x\by)^4 - (y\bx)^4 \right)
\big( (x\by)^4 +
(y\bx)^4 - 34\; (x\bx y\by)^2\cr
& + 12\; x\bx y\by (x\bx + y\by)^2 - (x\bx +
y\by)^4 \big)\cr }
\eqn\twentynine
$$
satisfying the relation
$$
Z^2 = Y\left(X^3 - 27\; Y^2 + W^2 (18\; X Y + X^2 W^2 + 16\; Y W^4)\right)
\eqn\thirty
$$
The locus of singular points is given by the two intersecting curves
$$
\eqalign{
& Z=0,\qquad X=-W^4, \qquad Y=0\cr
& Z=0, \qquad X = -{4\over3}\;W^4,\qquad Y = -{4\over 27}\;W^6 \cr }
\eqn\thirtya
$$

The $E_8$ ground ring is in principle obtained in the same way, but it
turns out to be rather involved and we will not write out the explicit
expressions here.

Thus (except for the $E_8$ case) we have explicitly constructed the full
non-chiral ground rings and their associated varieties at each of the
$A$-$D$-$E$ points. In each case, the singularities are described by a
pair of curves in \RR$^4$ intersecting at the origin, except at the
$SU(2)$ point where the only singular point is the origin and the variety
is a cone.

\section{\bf Non-chiral Ground Rings at Arbitrary Radii}

We turn now to the case of points in the moduli space of the $c=1$ free
boson which cannot be obtained by quotienting with a discrete subgroup of
$SU(2)$. This analysis proceeds via a direct study of allowed momenta at
various radii in $c=1$ CFT, and provides in particular a re-derivation of
the results presented above (for the $A_n$ and $D_n$ cases) directly from
conformal field theory.

The first case is that of circle compactification at rational radius in
units of ${1\over\sqrt2}$. Let $R= {p/q\over\sqrt2}$ with $(p,q)$ coprime
positive integers. The ground ring is obtained by examining the
momenta which are allowed at the given radius. The ground ring
elements $\Osn\bOsn$ at the $SU(2)$ point have left and right matter
momenta
$$
(p_L, p_R) = \sqrt{2}\;(n, n')
\eqn\thirtyone
$$
where $1-s \le n,n' \le s-1$ and $n,n'$ are both integer or half-integer.
Of these, at rational radii we must keep only
those operators whose matter momenta are given by
$$
(p_L, p_R) =
\left({M\over 2R} + NR, {M\over 2R} - NR\right) =
{1\over\sqrt2}\left({Mq\over
p}+{Np\over q},{Mq\over p}-{Np\over q}\right)
\eqn\thirtytwo
$$
which determines all the possible local vertex operators at the given
radius, in terms of two independent integers $M$ and $N$.

It follows that
$$
\eqalign{
n + n' &= {Mq\over p}\cr
n - n' &= {Np\over q}\cr  }
\eqn\thirtythree
$$
Since the left hand sides of both expressions above are integers, it
follows that we have ground ring elements only if $M,N$ are multiples of
$p,q$ respectively, and in that case we find the constraint that
$$
\eqalign{
n + n' &= 0~~{\rm mod}~q \cr
n - n' &= 0~~{\rm mod}~p \cr }
\eqn\thirtyfour
$$

Imposing this requirement is clearly equivalent to keeping polynomials in
the $SU(2)$ ground ring generators which are invariant under under the
$Z_p\otimes Z_q$ action
$$
\eqalign{
&x\bx\rightarrow e^{\displaystyle 2\pi i/q} ~x\bx \qquad
y\by\rightarrow  e^{\displaystyle -2\pi i/q} ~y\by \cr
&x\by\rightarrow e^{\displaystyle 2\pi i/p} ~x\by \qquad
y\bx\rightarrow e^{\displaystyle -2\pi i/p} ~y\bx \cr}
\eqn\thirtyfive
$$
This in turn can be obtained by defining the following action on the
chiral generators:
$$
\eqalign{
&x\rightarrow e^{\displaystyle i\pi/p} e^{\displaystyle i\pi/q}~x \qquad
\bx\rightarrow e^{\displaystyle -i\pi/p} e^{\displaystyle i\pi/q}~\bx \cr
& y\rightarrow e^{\displaystyle -i\pi/p} e^{\displaystyle -i\pi/q}~y \qquad
\by\rightarrow e^{\displaystyle i\pi/p} e^{\displaystyle -i\pi/q}~\by \cr}
\eqn\thirtysix
$$
It is easy to check that in the non-chiral ring there are five basic
invariants:
$$ V = (x\bx)^q \qquad W = (y\by)^q \qquad X = (x\by)^p \qquad
Y = (y\bx)^p \qquad Z = x\bx y\by
\eqn\thirtyseven
$$
satisfying the two relations
$$
\eqalign{
VW &= Z^q\cr
XY &= Z^p\cr }
\eqn\thirtyeight
$$

Thus we have a 3-variety given by two polynomial relations in \RR$^5$, as
long as $p,q \ne 1$. This is qualitatively different from the $A$-$D$-$E$
cases. Note also that, as expected, the generator $\half(x\bx + y\by)$ is
absent since it is invariant only under $SU(2)$ transformations for which
$(\bx,\by)$ transform as the complex conjugates of $(x,y)$. The
transformation in Eq.\thirtysix\ does not have this property, hence it
does not preserve the interpretation that we have been using above of
$(\bx, \by)$ as the complex conjugate coordinates of $(x,y)$. That
interpretation was necessary only to make contact with the Kleinian
theory, which we cannot do anyway at rational radii other than integers.

In the integer-radius case $q=1$, the first relation above can be solved
for $Z$, and we then have four generators satisfying one relation. A
trivial redefinition leads to Eq.\twentyfour. In this case, $(\bx,\by)$
transform as the complex conjugates of $(x,y)$. The other case, $p=1$,
gives the dual result with the role of electric and magnetic operators
interchanged.

For radii which are irrational multiples of ${1\over\sqrt2}$ on the circle
line, the only degenerate fields in the matter theory are those with zero
momentum. It follows that the only generator of the ground ring is $x\bx
y\by$, and it satisfies no relation. Thus the ground ring here is the free
polynomial ring on one generator.

On the orbifold line, at rational radius, we must mod out by the
transformation Eq.\thirtysix\ as well as the orbifold symmetry
$x\bx \leftrightarrow $, $x\by \leftrightarrow y\bx$.
This gives us the set of four generators
$$
\eqalign{
W &= \half\left( (x\bx)^q + (y\by)^q \right)\cr
X &= \half\left( (x\by)^p + (y\bx)^p \right)\cr
Y &= {1\over 4}\left((x\bx)^q - (y\by)^q \right)\left((x\by)^p -
(y\bx)^p\right) \cr
Z &= x\bx y\by\cr  }
\eqn\thirtyfive
$$
with the relation
$$
Y^2 = (W^2 - Z^q)(X^2 - Z^p)
\eqn\thirtysix
$$
which is a nice generalization of Eq.\twentysix, reducing to that equation
when $q=1$.

At irrational radius on the orbifold line, the situation is the same as
for irrational radius on the circle. The ground ring has a single
generator $x\bx y\by$ with no relation.

Finally, we consider infinite radius. This has been studied in\WITTEN,
where the ground ring was found to be generated by $x\bx$ and $y\by$ with
no relation. This can be obtained as a special case of Eq.\twentythree\
above in the limit $n\rightarrow\infty$, for which the generators $X$ and
$Y$ disappear (consistent with the fact that magnetic operators cannot
exist for a noncompact boson) and we are left with $W$ and $Z$ of that
equation with no relation between them.

This completes the analysis of ground rings for $c=1$ string theory at all
points in the moduli space of the $c=1$ conformal field theory.

\section{\bf Symmetries of $c=1$ Strings}

It is now understood\WITTEN\WITZWIE\ that the unbroken symmetries
of $c=1$ string theory are the volume preserving diffeomorphisms of the
variety defined by the ground ring. At the $SU(2)$ point, this is a
quadric cone in \RR$^4$. The generators of unbroken symmetry
transformations have two descriptions, related by descent equations. In
one description, appropriate to string field theory, they are local
operators in the BRS cohomology of ghost number 1. These are listed in
Eq.\five\ above, where the index $s$ is the $SU(2)$ isospin and takes all
positive integer and half-integer values, while $n,n'$ vary from $-s$ to
$s$ and $1-s$ to $s-1$ respectively. The other description (following from
descent equations) gives the symmetry generators as integrals of conserved
spin-1 currents, but we will not need that here.

Now we may ask which of the symmetry generators survive when we pass to
other points in the $c=1$ CFT moduli space, particularly the $A$-$D$-$E$
points. It has been shown at the $SU(2)$ point\WITTEN\WITZWIE, that
the symmetry generators can be thought of as polynomial vector fields
acting on the ground ring.  For example, the
generators in Eq.\five\ have the description
$$
Y^+_{s,n}(z) \bOsmn(\bz)~
\sim ~H_{s-1, n, n'}(x, y, \bx, \by)
\left( (s-n)x \delx - (s+n) y\dely \right)
\eqn\thirtyseven
$$
where
$$
H_{s,n,n'}\equiv x^{s+n} y^{s-n} \bx^{s+n'} \by^{s-n'}
\eqn\thirtyeight
$$
Let us first re-write these in terms of the variables $a_i$ defined in
Eq.\onea. Since these variables are constrained by the relation\two, there
is no unique representation, but one finds that all polynomial vector
fields of the form
$$
{a_1}^{\alpha_1} {a_2}^{\alpha_2} {a_3}^{\alpha_3} {a_4}^{\alpha_4}
\left( (1 + \alpha_2 + \alpha_4)\left(a_1 \delaone + a_3\delathree\right)
- (1 + \alpha_1 + \alpha_3)\left(a_2\delatwo + a_4\delafour\right) \right)
\eqn\thirtynine
$$
modulo the relation $a_1 a_2 = a_3 a_4$, are exactly the same set as those
in Eq.\thirtyeight. (The $\alpha_i$ are arbitrary non-negative integers.)

More generally, we construct the vector fields
$$
\prod_{i=1}^4 a_i^{\alpha_i} \sum_{i=1}^4 \left(c_i a_i\delai\right)
\eqn\forty
$$
with two requirements: one is that they have vanishing divergence, so that
they are volume-preserving, and the other is that the coefficients $c_i$
(which depend on the $\alpha_i$) should be invariant under the
transformation
$$
\alpha_1\to\alpha_1 + 1\quad\alpha_2\to\alpha_2 + 1\quad
\alpha_3\to\alpha_3 - 1\quad\alpha_4\to\alpha_4 - 1
\eqn\fortyone
$$
It is easy to see that these are all the symmetries of the type in
Eq.\five\ and their conjugates. The ``new'' symmetries in Eq.\fiveaa\
cannot be represented in this way since they act in the 4-dimensional
space \CC$^2$ in such a way as to be trivial after projection to the
$U(1)$ quotient, the true (3-dimensional) ground ring.

With this description of the symmetries at the $SU(2)$ point, it is
straightforward to generalise them to the $A$-$D$-$E$ points discussed
above. For each case, we define polynomial vector fields
$$
X^{\alpha_X} Y^{\alpha_Y} Z^{\alpha_Z} W^{\alpha_W}
\left(c_X X\delX + c_Y Y\delY + c_Z Z\delZ + c_W W\delW\right)
\eqn\fortytwo
$$
and subject them to the divergence-free condition, along with the
polynomial relation between $X,Y,Z,W$ defining the ground ring. This gives
all the residual symmetries of the type of Eq.\five\ at the relevant
point.

It is straightforward to carry this out explicitly at the $A_n$ points.
Starting with the generators $X,Y,Z,W$ defined in Eq.\twentythree, we
define the linear combinations $\tW = W + Z$, $\tZ = W - Z$. The subset of
symmetry generators of type $Y^+\overline{{\cal O}}$ which survive at the
$A_n$ point is given by
$$
\eqalign{
\tW^{\alpha_\tW} \tZ^{\alpha_\tZ} X^{\alpha_X} Y^{\alpha_Y}
&\bigg( (1 + \alpha_{\tZ} + n\alpha_Y)\left(\tW\deltW + n X\delX\right)
\cr
& - (1 + \alpha_{\tW} + n\alpha_X)\left(\tZ\deltZ + n Y\delY\right)\bigg)\cr}
\eqn\fortythree
$$

Thus our classification of ground rings at the $A$-$D$-$E$ points in the
$c=1$ moduli space leads naturally to a description of the unbroken
symmetry generators at these points. Symmetries of the type Eq.\five\
turn out to be the $H$-invariant polynomial vector fields on the associated
Kleinian singularities, which preserve the holomorphic volume form.
Equivalently they are the volume-preserving vector fields on the
three-variety \CC$^2/(\Gamma\otimes U(1))$.

\section{\bf HyperK\"ahler Manifolds and the Moduli of $c=1$ String Theory}

In this section we attempt to explore the physical consequences of the
$A$-$D$-$E$ classification of ground rings of the $c=1$ string via
Kleinian singularities.

In Ref.\WITZWIE, (see also \KLEBAPOLY), an effective action for all the
discrete modes of the $c=1$ string at the $SU(2)$ point was written down.
This action is built out of a closed 2-form $F=dA$ and a scalar $\sigma$
which are functions on
\CC$^2$, with the action
$$
\int_{C^2} \sigma\; F\wedge F
\eqn\fiftyone
$$
There is a constraint that $F$ should be a polynomial 2-form, representing
the moduli of the theory at the $SU(2)$ point with the positive Liouville
dressing, while $\sigma$ is some derivative of a $\delta$-function at the
origin of \CC$^2$. Below, we will try to explore what sort of solutions this
action has if we relax this constraint.

Recall that both $F$ and $\sigma$ should respect the left-right matching
of Liouville momenta, since the Liouville field is noncompact. This means
that they should be invariant under the vector field $H$ defined in
Eq.\twentytwo. Equivalently, they should be expressible in terms of the
$a_i$, on which this constraint is automatically implemented. This
constraint implies that the 4-dimensional action written above projects
down to a 3-dimensional action
$$
\int_{C^2/U(1)} \sigma\; du\wedge da
\eqn\fiftytwo
$$
where the original 4-dimensional gauge field $A$ is dimensionally reduced
to a scalar $u$ and a 3-dimensional 1-form $a$, and the action lives on
the quotient of \CC$^2$ by the $U(1)$ transformation defined in
Eq.\twentyone.

Now the results that we have discussed in this paper provide us with the
effective action at any other of the $A$-$D$-$E$ points in the moduli
space. It is just
$$
\int_{C^2/\Gamma} \sigma\; F\wedge F = \int_{C^2/(\Gamma\otimes U(1))}
\sigma\; du\wedge da
\eqn\fiftythree
$$
where the 4-dimensional space on the left-hand side is precisely the
Kleinian singularity for the relevant subgroup $\Gamma$, while the
3-dimensional space on the right is defined by the true non-chiral ground
ring.

The equation of motion from this action,
$$
F\wedge F =0,\qquad dF=0
\eqn\fiftyfour
$$
is to be solved on a non-trivial space, unlike at the $SU(2)$ point. As a
result, if the topology of this space is nontrivial, we should
expect to find solutions for $F$ which are closed but not exact. (At the
$SU(2)$ point, the solutions of this equation are given by the (old and
new) moduli of Ref.\WITZWIE, all of which are closed since they are
annihilated by $b_0^-$, but also exact since they can be expressed as
$b_0^-$ on some other operator. Hence they are trivial in the
$d$-cohomology, which is expected since \CC$^2$ is a topologically trivial
manifold.)

This equation is well-known and has some very interesting solutions
(for a recent review, see Ref.\TAKASAKI). These are, however,
not generally $H$-invariant. Indeed, on a general
4-manifold, one can write a class of solutions of the form
$$
F = \omega + i\lambda G + \lambda^2 {\bar\omega}
\eqn\fiftyfive
$$
where, with respect to some complex coordinates on this 4-manifold, $\omega$
and ${\bar\omega}$ are closed (2,0) and (0,2) forms, $G$ is a closed (1,1)
form, and $\lambda$ is a parameter. The equation $F\wedge F=0$ turns into
$$
G\wedge G = 2 \omega\wedge {\bar\omega}
\eqn\fiftysix
$$
If we interpret $\omega$ as a holomorphic volume form and $G$ as a
$(1,1)$ K\"ahler form, then this equation defines a Ricci-flat
hyperK\"ahler manifold, which in 4 dimensions is equivalent to self-dual
gravity. Indeed, the above equation is the Plebanski equation\PLEBANSKI,
which has in particular been discussed in some detail in Ref.\OOGURIVAFA\
in the context of $N=2$ strings. We will see below that while the $(1,1)$
part of $F$ may be chosen consistent with $H$-invariance, the $(2,0)$ part,
which is associated to the volume form, is not $H$-invariant.

To study this in more detail, note that on a complex manifold the
K\"ahler form $G$ can be written
$$
G = \del\delbar\Phi(z^i,\bz^i)
\eqn\fiftyseven
$$
where $z^i$ are complex coordinates on the manifold and
$\Phi(z^i,\bz^i)$ is the K\"ahler potential. Inserting this into
Eq.\fiftysix\ leads to
$$
\del\delbar\Phi\wedge \del\delbar\Phi = 2\omega\wedge{\bar\omega}
\eqn\fiftyeight
$$
Thus the dynamics of this system can be defined in terms of a single
scalar function on the manifold (although this is not globally defined,
but changes across coordinate patches).

In $c=1$ string theory at the $SU(2)$ point, we can use the ground ring
generators $x,y$ as the complex coordinates. The conventional moduli
have the following description as 2-forms:
$$
Y^+_{s,n}\bY{}^+_{s,n'}~\sim~dh_{s,n}\wedge d\bh_{s,n'}
\eqn\fiftyeighta
$$
where $h_{s,n}$ is defined in Eq.\new\ and $\bh_{s,n'}$ is its
antiholomorphic counterpart. Now the above expression can be written as a
closed $(1,1)$ form of the type of Eq.\fiftyseven\ above, where
$$
\Phi(x,y,\bx,\by)= H_{s,n,n'} \equiv h_{s,n}\bh_{s,n'}
\eqn\fiftyeightb
$$
A simple example is
$$
\Phi(x,y,\bx,\by) = x\bx + y\by
\eqn\fiftynine
$$
The corresponding modulus is
$$
Y^+_{\half,\half}\bY{}^+_{\half,\half} +
Y^+_{\half,-\half}\bY{}^+_{\half,-\half}~\sim~
dx\wedge d\bx + dy\wedge d\by
\eqn\sixty
$$
which is the K\"ahler form for the flat metric on \CC$^2$.
In fact, all the moduli of this type are exact at the $SU(2)$ point, so
none of them can be the K\"ahler form for a nontrivial metric.

We note in passing that the above modulus, corresponding to the flat
K\"ahler metric, exists at all the $A$-$D$-$E$ points in the moduli
space, but nowhere else. Indeed, the K\"ahler potential $\Phi$ in
Eq.\fiftynine\ is precisely the generator $W$ of the polynomial ring at
all these points.

Consider now the ``new moduli'' of the type
$$
\eqalign{
(a+\ba)\oct_{s-1,n}\bY{}^+_{s,n'}~\sim~
&H_{s-1,n,n'}\big(\bx\by\; dx\wedge dy + {s+n'\over 2s} y\by\; dx\wedge d\bx
+ {s-n'\over 2s} y\bx\; dx\wedge d\by \cr
&- {s+n'\over 2s} x\by\; dy\wedge d\bx
- {s-n'\over 2s} x\bx\; dy\wedge d\by \big) \cr  }
\eqn\sixtya
$$
discovered in\WITZWIE. These are the sum of a $(2,0)$ and a $(1,1)$ piece.
The sum is closed (and exact) but the individual pieces are not. Thus it
is not possible to identify any of the new moduli with the closed
$(2,0)$ piece $\omega$ in Eq.\fiftyfive\ describing the volume form
$dx\wedge dy$. This is anyhow obvious since $dx\wedge dy$ is not an
$H$-invariant form.

One may ask which new modulus would correspond to this volume form if we
relaxed the condition of $H$-invariance. It was noted in\WITZWIE\ that in
the chiral cohomology, there is a special operator
$$
{\tilde\iota} =
c\del c e^{\displaystyle\sqrt{2}\phi}
\eqn\sixtyone
$$
which plays the role of $a\cdot Y^+_{0,0}$, and is represented as a
constant bivector on the ground ring, or equivalently a constant function
in the dual (form) representation. Starting with the antiholomorphic
counterpart of this, there is no ground ring element in the holomorphic
sector which can be paired with it to form an $H$-invariant modulus, since
the minimum isospin $s$ for new moduli is 1, from Eq.\sixtya. However, if
Liouville momentum matching were not a constraint, we could have chosen an
element like $(a + \ba)\oct_{0,0}\bY{}^+_{0,0}$ which is precisely
$dx\wedge dy$. This shows explicitly that, to obtain the geometric
interpretation of Eq.\fiftyfive, one must relax the $H$-invariance
condition, which corresponds to taking solutions of the four-dimensional
action which do not descend to solutions of the three-dimensional one.

Thus, even at the $SU(2)$ point, solutions of the equation
of motion \fiftyfour\ of the type \fiftyfive, which have a geometric
interpretation, contain an $H$-noninvariant piece $\omega$.
We have exhibited a solution of this type, which corresponds to the
(trivial) hyperK\"ahler geometry of flat \CC$^2$. At the other $A$-$D$-$E$
points, solutions exist which correspond to non-trivial hyperK\"ahler
geometry, as we discuss below.

It is tempting to think of $H$-invariant solutions as time-independent
solutions of Eq.\fiftyfour, while the nontrivial hyperK\"ahler
manifolds which also solve the equations of motion could be thought of as
time-dependent solutions, rather like instantons.

Recall now that the Abelian gauge field strength $F$ appearing in the
$c=1$ string effective action is really a collection of the degrees of
freedom corresponding to all the discrete states in the theory. The
discussion above suggests that the dynamics of these modes is that of {\it
self-dual gravity in a target space defined by the ground ring of the
theory}, which is four-dimensional in the sense discussed above.

Now to explicitly find nontrivial solutions of the equation of motion for
the action\fiftythree\ for a given subgroup $\Gamma$ of $SU(2)$, we need a
class of hyperK\"ahler metrics consistent with the topology of
\CC$^2/\Gamma$. Remarkably, this problem has been completely solved and
leads to a very beautiful result\HITCHIN\KRONHEIMER. To every Kleinian
variety \CC$^2/\Gamma$, there corresponds a finite-parameter family of
hyperK\"ahler manifolds obtained by resolving the singularity at the
origin. These manifolds are precisely the gravitational
instantons\HAWKING. The first in the series (corresponding to
$\Gamma=\cyc_2$) is the Eguchi-Hanson instanton\EGUCHI, while
$\Gamma=\cyc_n$ is in correspondence with the multi-center instanton
metrics of Gibbons and Hawking\GIBBONS\HITCHIN. More general instanton
solutions are known corresponding to blowing up the singularities of the
$D$ and $E$-series Kleinian varieties\KRONHEIMER\ (see also \DANCER).
These instantons are all asymptotically locally Euclidean (ALE) spaces,
which means that at infinity their topology is that of $S^3/\Gamma$.
In each such case there is a finite-parameter moduli space of solutions
(this is $3n-6$-dimensional in the $\cyc_n$ case, for example).

Thus we conclude that at each of the $A$-$D$-$E$ points in the moduli
space of $c=1$ string theory, associated with the discrete subgroups
$\Gamma$ of $SU(2)$, there is a finite-parameter family of nontrivial
classical solutions of the four-dimensional string action corresponding
precisely to the blown up version of the Kleinian variety \CC$^2/\Gamma$,
which plays the role of the ``target space'' of the theory, and these
solutions are the various hyperK\"ahler metrics consistent with the
topology of this space.

The principal problem with this interpretation, as we have noted above, is
that it appears to be inconsistent with $H$-invariance, which itself is a
consequence of the noncompactness of the Liouville field.  We need a
better understanding of the physical meaning of $H$-invariance, and possibly
a compactified Liouville field, from the ground-ring target-space point of
view.

\section{\bf Conclusions}

We have shown that the $A$-$D$-$E$ classification of ``special points'' in
the moduli space of the $c=1$ free boson conformal field theory results in
an analogous classification of ground rings at these points when the CFT
is coupled to 2D gravity to form a string theory. The ground rings and
their corresponding algebraic varieties so obtained are closely related to
the singular varieties studied by Klein in connection with the discrete
subgroups of $SU(2)$. The unbroken symmetries of the theory at each such
point form the algebra of volume-preserving diffeomorphisms of these
varieties.

We have studied the theory at zero cosmological constant. It would be
interesting to see how the polynomial ground rings at various points in
the moduli space are deformed under cosmological perturbation. At the
$SU(2)$ point, the singularity at the origin is smoothened out under this
perturbation\WITTEN\ (see also Refs.\con\BARBON\KACHRU\noc). However, at
the other points we have shown that the singularities are not isolated
but lie along curves, so it is no longer clear that a cosmological
perturbation will produce a smooth manifold.

Perhaps the most interesting consequence of our results appears in the
relation of Kleinian singularities to hyperK\"ahler surfaces, which are
solutions of self-dual gravity in four dimensions. Mathematically,
resolving the singularity at the origin produces families of hyperK\"ahler
metrics for each Kleinian variety. These metrics are known in
four-dimensional Einstein gravity, where they appear as gravitational
instantons, and are interpreted (as in Yang-Mills theory) as tunneling
solutions between distinct vacua of the theory. It is clearly important to
understand whether they play an analogous or different role in the context
of string theory. In any case it is very suggestive that the collection of
discrete states in $c=1$ string theory has an effective action (in the
ground ring space) which has gravity-like solutions, as this seems to be
quite different from the role normally played by gravitation in the
context of strings. A dynamical understanding of this phenomenon would
give us some clues about the underlying structure of string theory.

\noindent{\bf Acknowledgements}\hfill\break
We would like to thank Sumit Das, Kirti Joshi, Gautam Mandal, Sudipta
Mukherji, Kapil Paranjape, T. Ramadas, Ashoke Sen, Anirvan Sengupta and V.
Srinivas for helpful discussions.

\refout
\end